%% file: 0_main.tex
\documentclass[conference]{IEEEtran}
\usepackage[utf8]{inputenc}
\usepackage{url}
\usepackage{listings}
\usepackage{multirow}
\usepackage{xcolor}
\usepackage{textcomp}
\usepackage{color}
\usepackage{graphicx}
\usepackage{caption}
\usepackage{subcaption}
\usepackage{balance}

\input{util}

\begin{document}

\title{Are We Ready to Embrace Generative AI \\ for Software Q\&A?}

\author{
    \IEEEauthorblockN{Bowen Xu\IEEEauthorrefmark{1}\IEEEauthorrefmark{2}, Thanh-Dat Nguyen\IEEEauthorrefmark{3}, Thanh Le-Cong\IEEEauthorrefmark{3}, Thong Hoang\IEEEauthorrefmark{4}, Jiakun Liu\IEEEauthorrefmark{2}, \\Kisub Kim\IEEEauthorrefmark{2}, Chen Gong\IEEEauthorrefmark{5}, Changan Niu\IEEEauthorrefmark{6}, Chenyu Wang\IEEEauthorrefmark{2}, Bach Le\IEEEauthorrefmark{3}, David Lo\IEEEauthorrefmark{2}}

    \IEEEauthorblockA{\IEEEauthorrefmark{1}\textit{North Carolina State University, USA}
    \\bxu22@ncsu.edu}
    \IEEEauthorblockA{\IEEEauthorrefmark{2}\textit{Singapore Management University, Singapore}
    \\\{bowenxu.2017, jkliu, kisubkim, chenyuwang, davidlo\}@smu.edu.sg}
    \IEEEauthorblockA{\IEEEauthorrefmark{3}\textit{University of Melbourne, Austrialia}\\\{thanhdatn, congthanh.le\}@student.unimelb.edu.au, bach.le@unimelb.edu.au}

    \IEEEauthorblockA{\IEEEauthorrefmark{4}\textit{CSIRO's Data61, Australia} \qquad \qquad
   \IEEEauthorrefmark{5}\textit{University of Virginia, USA}     \qquad\IEEEauthorrefmark{6}\textit{Nanjing University, China}\\james.hoang@data61.csiro.au \qquad  \qquad  fzv6en@virginia.edu     \qquad \qquad \qquad niu.ca@outlook.com} \qquad

    \vspace{-1cm}
}

\maketitle

\begin{abstract}
Stack Overflow, the world's largest software Q\&A (SQA) website, is facing a significant traffic drop due to the emergence of generative AI techniques.
\gpt{} is banned by \so{} after only 6 days from its release.
The main reason provided by the official \so{} is that the answers generated by \gpt{} are of low quality.
To verify this, we conduct a comparative evaluation of human-written and ChatGPT-generated answers. Our methodology employs both automatic comparison and a manual study.
Our results suggest that human-written and \gpt{}-generated answers are semantically similar, however, human-written answers outperform \gpt{}-generated ones consistently across multiple aspects, specifically by 10\% on the overall score.
We release the data, analysis scripts, and detailed results at \url{https://github.com/maxxbw54/GAI4SQA}.
\end{abstract}

\vspace{2mm}
\section{Introduction}
\label{sec:intro}
\input{1_Introduction}

% \vspace{2mm}
\section{Data Preparation}
\input{2_Methodology}
\section{RQ1: Automatic Comparison}
\label{sec:rq1}
\input{3_RQ1-AutoCompare}

% \vspace{2mm}
\section{RQ2: Manual Comparison}
\label{sec:rq2}
\input{4_RQ2-UserStudy}

\vspace{2mm}
\section{Threats to Validity}
\input{5_Threats}

% \section{Lessons Learned}
% \input{6_Lessons_learned}

\vspace{2mm}
\section{Related Work} \label{sec:related_work}

\input{7_related_work}

\vspace{2mm}
\section{Conclusion}
\input{8_Conclusion}

% \section{Supplementary Material}
% \input{9_Supplementary}

% \section*{Acknowledgment}
% This research / project is supported by the National Research Foundation, Singapore, under its Industry Alignment Fund – Pre-positioning (IAF-PP) Funding Initiative. Any opinions, findings and conclusions or recommendations expressed in this material are those of the author(s) and do not reflect the views of National Research Foundation, Singapore.

\balance

\bibliographystyle{IEEEtran}
\bibliography{ref}

\end{document}

%% file: util.tex
\newcommand{\so}[0]{Stack Overflow}
\newcommand{\gpt}[0]{ChatGPT}

\usepackage{xcolor}
\usepackage{enumitem}
\usepackage{amsmath}

\usepackage[strict]{changepage}
\usepackage{framed}

\definecolor{formalshade}{rgb}{0.95,0.95,1}

\newenvironment{formal}{%
  \MakeFramed{\advance\hsize-\width\FrameRestore}%
  \noindent\hspace{-4.55pt}% disable indenting first paragraph
  \begin{adjustwidth}{}{7pt}%
  \vspace{2pt}\vspace{2pt}%
}
{%
  \vspace{2pt}\end{adjustwidth}\endMakeFramed%
}

\usepackage{booktabs}
\usepackage{xurl}

\usepackage[numbers,sort&compress]{natbib}

%% file: 1_Introduction.tex
On November 30, 2022, OpenAI, a world-class AI company, launched an artificial intelligence chatbot named \gpt{}~\cite{chatgpt}. Since then, it has rapidly become a widely used tool because of its impressive ability to produce responses on various tasks. 
% According to a UBS study,\footnote{\url{https://www.reuters.com/technology/chatgpt-sets-record-fastest-growing-user-base-analyst-note-2023-02-01/}} \gpt{} has reached 100 million users just two months after launching, making it the fastest-growing consumer application in history. 
In just two months after its release, ChatGPT reportedly reached 100 million users, making it the fastest-growing consumer application in history~\cite{chatgptuser}.
\gpt{} interacts with users as a chatbot; therefore, the most human-comparable usage of \gpt{} is for question and answering. 
In the context of software engineering (SE), \gpt{} has already been widely used by programmers to answer technical queries~\cite{chatgpt4sqa}.
% After only 6 days from \gpt{} was released, on 05 December 2022, it is temporarily banned by \so{}\footnote{\url{https://meta.stackoverflow.com/questions/421831/temporary-policy-chatgpt-is-banned}}, the biggest Software Q\&A crowdsourcing platform with 24 million technical questions and 35 million answers\footnote{\url{https://stackexchange.com/sites?view=list\#traffic}}. The reason provided by \so{} officially quoted:
However, the application was banned by \so{}~\cite{sobanchatgpt}, the largest software Q\&A crowdsourcing platform, only after six days from its release. We quote the reason provided by Stack Overflow as follows:

\begin{formal}
``Overall, because the average rate of getting correct answers from \gpt{} is too low, the posting of answers created by \gpt{} is substantially harmful to the site and to users who are asking and looking for correct answers.''
\end{formal}

Despite the above claim from Stack Overflow, there remains no clear empirical evidence on the overall quality of ChatGPT-generated responses as compared to human-written ones on software question answering (SQA). In this booming era of AI-powered chatbots, traffic to OpenAI's ChatGPT has been growing exponentially, while traditional Q\&A site such as Stack Overflow has been experiencing a steady decline~\cite{sotraffic}. Specifically, traffic to Stack Overflow was down by 6\% every month in January 2022 on a year-over-year basis and was down 13.9\% in March 2022~\cite{sodecline}. This phenomenon, however, is concerning due to the lack of empirical evidence on a comparative study on human-written vs AI-generated responses. The empirical evidence is much needed to ensure a balanced and robust development in the field of SQA.
In this work, we investigate the following research questions:

% \vspace{2mm}
% \noindent\fbox{\begin{minipage}{.95\columnwidth}
% \textbf{Hypothesis} \emph{For SQA, ChatGPT-generated answers are semantically similar to human-written answers; however, ChatGPT-generated answers have lower quality compared to human-written answers.}
% \end{minipage}}
% \vspace{2mm}\\
% To verify this hypothesis, we present two research questions:

\begin{itemize}
    \item \textbf{RQ1:} \emph{What are the characteristics of \gpt{}-generated and human-written answers?}
    \item \textbf{RQ2:} \emph{From the human user perspective, how good are the \gpt{}-generated answers?}
\end{itemize}

% \bowen{one sentence to summarize the methodology and then present the key findings of RQ1 and RQ2, and overall...}

% We find that (1) surprisingly, the length of human answers is longer than \gpt{} answers on average but the number is less; (2) overall, \gpt{}-generated and human-written answers are 86\% similar to each other; and (3) human and \gpt{} have similar opinions on whether code snippet needs to be provided in the answers or not.
% To answer RQ2, we conduct a user study to evaluate the human and \gpt{}-generated answers. Inspired by~\ref{}, we design questionnaires and ask participants to compare human and \gpt{}-generated answers from many different perspectives, such as relevance, correctness, etc. Moreover, we also ask participants to see if any factual error exists in the \gpt{}-generated answers since \emph{hallucination}\footnote{\url{https://cdn.openai.com/papers/gpt-4.pdf}} has been heavily complained about by its users and heatedly discussed by the public.
% We find that the human answers are still better than \gpt{} answers on all the considered aspects.
% However, human answers only have a small lead on some of the aspects, such as diversity, readability, and clarity.

% We also ask participants to see if any factual errors exist in the \gpt{}-generated answers since hallucinations have been heavily complained about by its users and heatedly discussed by the public.
% We find that human-written answers are still better than \gpt{}-generated answers. However, the human-written answers only hold a slight lead in some aspects, such as diversity, readability, and clarity.
%

We answer RQ1 and RQ2 by considering automatic metrics and conducting a manual study, respectively.
For RQ1, we find that (1) the average length of answers may not always be consistent
with the binary result of which answer is longer than the other, (2) the semantics of
human and ChatGPT answers are close to each other, (3) humans and ChatGPT have significantly different opinions on whether the questions should be answered with code snippets or not.
For RQ2, we find that (1) human-written answers are still better than \gpt{}-generated answers from 6 aspects (Correctness, Usefulness, Diversity, Readability, Clarity,
and Conciseness), (2) \gpt{}-generated answers can fully address only 52\% of the questions while human answers can fully address 84\%, (3) 27\% of \gpt{}-generated answers carry factual errors while only 2\% human answers have factual errors.
Overall, we accept our hypothesis with both quantitative and qualitative evidence, i.e., \gpt{}-generated answers are semantically similar to human answers; however, they are of lower quality.

Our research result can be useful for future SQA from multiple perspectives, as we describe it in more detail throughout the paper.
We find that human users can easily distinguish human-written and \gpt{}-generated answers. It mitigates the risk of adopting \gpt{} for SQA.
Moreover, we find that \gpt{} tends to be conservative in answering different questions differently.
We also find that \gpt{} and humans have different opinions on whether a question should be answered with code or not.
Nevertheless, the overall quality of \gpt{}-generated answers is still fair and \gpt{} can immediately generate answers which is much faster than waiting for an acceptable answer on \so{} (more than 13 days on average). Therefore, our experimental results suggest that the following directions are worth further research: (1) how to improve the ChatGPT-generated answers, (2) how to design better interplay between humans and ChatGPT for better SQA.

% The main contributions of our paper are as follows:
% \bowen{the implication of our paper is...}

% \begin{itemize} [leftmargin=*]
%     \item To the best of our knowledge, we are the first to investigate the capability of \gpt{} specifically in SQA and analyze the characteristics of \gpt{}-generated and human-written answers.
%     \item Based on our results, we find that \gpt{}-generated answers are still of low quality as compared with human-written answers.
%     \item We summarize a list of potential research directions to boost automatic SQA for software practitioners.
% \end{itemize}

%% file: 2_Methodology.tex
% In this section, we present an overview of our framework.
In this section, we describe how we collect technical questions and answers generated by \gpt{} and humans.
% We introduce the evaluation methodology to estimate the quality of \gpt{}-generated vs. human-written answers.

% \subsection{Overview}\label{sec:overview}

% \begin{figure}[h]
%     \centering
% \includegraphics[width=\columnwidth]{figs/overview_1.pdf}
%     \caption{An overview of our methodology}
%     \label{fig:overall}
%     \vspace{-3mm}
% \end{figure}

% Figure~\ref{fig:overall} presents the framework of our methodology which consists of three main steps:
% Figure~\ref{fig:overall} illustrates the overview framework of our methodology with three main steps:

% \begin{enumerate} [leftmargin=*]
%     \item \emph{Technical question collection}. This step aims to build a sample set of techniques based on a collection of reasonable criteria.
%     \item \emph{Answer collection}. This step collects the corresponding \gpt{}-generated and human-written answers for each selected question.
%     \item \emph{Evaluation}. This step compares and evaluates the \gpt{}-generated and human-written answers based on automatic metrics and user study.
% \end{enumerate}
% Details of each step are explained in the following subsections.

\subsection{Technical Question Collection}\label{sec:rq1_data}

% We use the following criteria to initialize our search space of technical questions from \so{}.

Following previous works~\cite{calefato2018ask, ahasanuzzaman2016mining, baltadzhieva2015predicting}, we define some criteria to initialize our search space for technical questions on Stack Overflow: (1) As \gpt{} was trained based on the accumulated dataset from September 2021 onwards, we collect questions created in 2022 to mitigate data leak issue; (2) Each question must have an accepted answer; (3) Questions are tagged with a specific programming language. In this paper, we consider the questions tagged as ``Java'' and ``Python''; (4) Each question does not duplicate any other questions; (5) Questions have more than 5 upvotes. Intuitively, the questions with higher votes are more likely to be described clearly, (6) We select questions that do not include images as ChatGPT cannot process images.
% Due to the fact that the image recognition feature of GPT-4 is not accessible yet, we skip the questions with images.
% Table~\ref{tab:data_stats} presents the statistic of the collected questions.
% Based on the selection criteria, we collect 442 and 182 questions relate to Python and Java, respectively.
% Correspondingly, the average length, i.e., the numbers of words, of each Python and Java question are 198 and 225.
% Specifically, we also calculate the average number of tokens for each question since it can be considered as the input length to \gpt{}.
% On average, each Python and Java question contains 420 and 576 tokens.
% We find that the ratio between the number of words and tokens is 1:2.12 and 1:2.56 in Python and Java questions.
% Table~\ref{tab:data_stats} presents the statistics of the collected questions.
Based on the selection criteria, our dataset contains 442 and 182 questions related to Python and Java, respectively. The average length, in terms of the number of words, of each Python and Java question is 198 and 225, respectively. Correspondingly, each Python question contains 420 tokens, while each Java question includes 576 tokens.
% The ratio between the number of words and tokens is 1:2.12 and 1:2.56 in Python and Java, respectively.

% \usepackage{multirow}
% \begin{table}[h]
% % \vspace{-3mm}
% \centering
% \begin{tabular}{cccc}
% \toprule
% \multirow{2}{*}{PLs} & \multirow{2}{*}{\# Questions} & \multicolumn{2}{c}{For each question*} \\ \cline{3-4} 
%                     &                                  & Avg. \# Words        & Avg. \#  Tokens      \\ \hline
% Python              & 442                              &         198           &   420          \\
% Java                & 182                              &         225           &  576        \\ \bottomrule
% \end{tabular}
% \\
% $*$: Question Title and Body.
% \vspace{-2mm}
% \caption{Statistic of Collected Questions}
% \label{tab:data_stats}
% \vspace{-6mm}
% \end{table}

\subsection{Answer Collection}\label{sec:ans_collect}

% We consider the accepted answers in \so{} as human answers with high quality. And we use OpenAI API\footnote{\url{https://platform.openai.com/docs/api-reference/chat/create}} to query the model \emph{gpt-3.5-turbo} by using its default parameter setting and the following prompt.

We consider \emph{accepted answers} on Stack Overflow as these answers, written by humans, are often of high quality~\cite{soacceptedanswer}. We employ the OpenAI API~\cite{openaitextsim} to query the model \emph{gpt-3.5-turbo}.
We set the system prompt (i.e., the role of the \gpt{} in our task) to \emph{You are a software question and answer chatbot for programmers}.
And for the user prompt (i.e., the structured content of the question), we use a simple prompt \emph{Question title : [Title] [NEWLINE] Question body : [BODY]}.

%% file: 3_RQ1-AutoCompare.tex
% This section aims to answer the research question, to what extent, the \gpt{}-generated answers are different from human-written answers.
To answer RQ1, we employ a set of automatic metrics on the collected question and answer pairs.

\subsection{Metrics}

% For each technical question, we measure the text similarity between \gpt{}-generated and human-written answers by using two commonly used metrics.

\noindent \textbf{Length}.
We consider the \emph{length} of an answer as a proxy for measuring readability and conciseness. Thus, we empirically investigate the length of human-written and ChatGPT-generated answers.
Specifically, we calculate the length from two granularity levels: \emph{words} (appreciated by humans) and \emph{tokens} (considered by ChatGPT).
For each answer, we calculate the number of words by considering the space as the delimiter. The GPT family of models process text using tokens, which are common sequences of characters found in text. In this work, we calculate the number of tokens by using the OpenAI tokenizer~\cite{openaitokenizer}.

% \noindent \textbf{ROUGE score}. 
% % Following the existing extractive summarization approaches \cite{chengran2022answer}, we use ROUGE-N for automatic evaluation.
% % ROUGE-N is a standard automatic evaluation metric for summarization systems. It evaluates the n-gram overlapping between a generated summary and the ground-truth summaries.
% % ROUGE-N is defined as:
% ROUGE-N is a standard automatic evaluation metric for summarization systems~\cite{chengran2022answer}. The metric estimates the n-gram overlap between a generated summary and the ground-truth summaries. ROUGE-N is presented as follows:

% \begin{equation}
% \text{ROUGE-N}=\frac{\sum\limits_{S_{i}\in{S}}\sum\limits_{gram_{n}\in{S{{i}}}}Count_{match}(gram_{n})}{\sum\limits_{S_{i}\in{S}}\sum\limits_{gram_{n}\in{S{{i}}}}Count(gram_{n})}
% \end{equation}
% %
% where $S$ denotes the ground-truth summary, $n$ denotes the number of grams in a summary, $Count_{match}(gram_{n})$ and $gram_{n}$ denote the number of grams that are coexisting in the ground-truth summaries and generated summaries. 
% We apply ROUGE-N to evaluate the quality of the generated summaries against the ground-truth summaries. The higher the value of ROUGE-N, the better the performance of an approach. \jh{how can we use ROUGE-N to measure ChatGPT-generated vs. human-written answers?}

\noindent\textbf{Similarity}. We measure the similarity between human-written and \gpt{}-generated answers based on their embeddings. That is we map two pieces of text to a semantic vector space and then calculate their distance in the vector space to reflect their similarity. The closer they are, the more semantically similar they are. To implement this, we use the latest embedding model developed by OpenAI, i.e., \emph{text-embedding-ada-002}, and calculate the similarity by following the official instructions~\cite{openaiembeddings}.

\noindent\textbf{Code recommendation}. Considering that code snippets are commonly provided in the answers to technical questions. We investigate if there is a significant difference between humans and \gpt{} in determining whether a question needs to be answered with code or not.

\subsection{Experimental Result}

% \usepackage{multirow}
% \usepackage{graphicx}
\begin{table}[h]
\caption{Automatic Comparison between Human and \gpt{}-generated answers}
\label{tab:auto_compare}
\resizebox{\columnwidth}{!}{%
\begin{tabular}{cccccc} \toprule
PLs                     & Answer Type       & \begin{tabular}[c]{@{}c@{}}Avg. \# of\\ Words\end{tabular} & \begin{tabular}[c]{@{}c@{}}Avg. \# of\\ Tokens\end{tabular} & \begin{tabular}[c]{@{}c@{}}\# if longer\\ than the other\end{tabular} & Similarity            \\ \hline
\multirow{2}{*}{Python} & Human-written & 323              & 387               & 207                         & \multirow{2}{*}{0.86} \\
                        & \gpt{}-generated   & 172              & 236               & 235                         &                       \\ \hline
\multirow{2}{*}{Java}   & Human-written & 314              & 345               & 93                          & \multirow{2}{*}{0.86} \\
                        & \gpt{}-generated   & 173              & 219               & 89                          &                   \\   \bottomrule
\end{tabular}%
}
\end{table}

Table~\ref{tab:auto_compare} presents the results of RQ1. On the one hand, we surprisingly find that human-written answers carry more words and tokens than \gpt{}-generated answers. For both Python and Java questions, the human-written answers are around 1.8 times longer than \gpt{}-generated answers. On the other hand, among all the questions considered in this work, 8\% of their human-written answers are shorter than \gpt{}-generated answers. It indicates that \textbf{the average length of answers may not consistently align with the binary outcome of determining which answer is longer than the other}.
Moreover, based on the similarity calculated based on OpenAI text embedding model, we find that the similarity between human and \gpt{}-generated answers to both Python and Java questions only differ slightly. And \textbf{overall, the semantics of human and \gpt{}-generated answers are close to each other}.
% As shown in Figure~\ref{fig:python_code} and~\ref{fig:java_code}, the Venn diagrams present the proportions of the answers with code to Python and Java questions, respectively.
% Besides, we find that \gpt{} tends to generate an answer with code snippets for Java questions (55\%) than Python questions (35\%).
% Also, we find that for only 59\% of questions, both human and \gpt{} have the same opinion on whether the questions should be answered with code snippets or not.
Moreover, we calculate \emph{Cohen's kappa coefficient}~\cite{cohen1960coefficient} to measure the agreement between humans and \gpt{} on determining whether a question should be answered with code or not. We find that the kappa score is only 0.07 which corresponds to the \emph{slight agreement}. It indicates \textbf{humans and \gpt{} have significantly different opinions on whether the questions should be answered with code snippets or not}.
% , i.e., the overlap area in the Venn diagrams.

% \bowen{find the different and ask human labelers.}

% \begin{figure}[h]
%      \centering
%      \begin{subfigure}[b]{0.4\textwidth}
%          \centering
%     \includegraphics[width=\textwidth]{figs/python.pdf}
%          \caption{Ratio of Answers with Code to 442 Python Questions}
%          \label{fig:python_code}
%      \end{subfigure}
%      \hfill
%      \begin{subfigure}[b]{0.4\textwidth}
%          \centering
%     \includegraphics[width=\textwidth]{figs/java.pdf}
%          \caption{Ratio of Answers with Code to 182 Java Questions}
%          \label{fig:java_code}
%      \end{subfigure}
%         \caption{Percentage of Answers w/wo Code}
%         \label{fig:Answer_code}
% \end{figure}

%% file: 4_RQ2-UserStudy.tex
To answer RQ2, we conduct a manual analysis to evaluate \gpt{}-generated and human-written answers. 
% Specifically, we carefully design customized questionnaires for each of the questions and ask participants to compare \gpt{}-generated and human-written answers from various aspects.
% We aim to investigate how good or bad the \gpt{}-generated answers.
% We randomly sample a subset of 20 technical questions collected in Section~\ref{sec:rq1_data}.

% To have a deeper understanding of pros and cons of human and \gpt{} generated answers, we conduct a user study with real-world technical questions.

\subsection{Participants}
We design a customized questionnaire for each question and then distribute it to the participants. In total, we invite 7 participants.
For questions related to different programming languages, we assign them to participants who have at least 2 years of programming experience. We ask participants to skip questions for which one or more of these conditions are satisfied: (1) the participants think they are not knowledgeable enough to answer the questions, (2) the participants have read the given question and its corresponding answers before.
Moreover, we prohibit participants to search the original Stack Overflow post during the manual study.

\subsection{Questionnaire Design}
Inspired by~\cite{xu2020reinventing}, we randomly sort the answers generated by human and \gpt{} across different questionnaires.
Thus, our participants may not receive the answers in the same order.
Before asking questions, we first present the title and the body of the target technical question.
Then, we present an answer either generated by humans or \gpt{}.
For each answer, we ask the following questions:

\noindent \textbf{Q1.} \emph{How satisfied are you with the answer? (Required)}

\noindent Inspired by prior works~\cite{xu2017answerbot,yang2022answer}, we consider 6 aspects to measure the quality of an answer on a 5-point Likert scale (from 1: Very dissatisfied to 5: Very satisfied). The 6 aspects are: Correctness, Usefulness, Diversity, Readability, Clarity, and Conciseness. Besides, we also ask participants to measure the overall quality of the given answer.

\noindent \textbf{Q2.} \emph{Please explain your rate on the answer.}

\noindent Besides the scores in Q1, we encourage participants to provide further explanations for their ratings.

\noindent \textbf{Q3.} \emph{Do you think the answer correctly understands the question? (Required)}

\noindent $\bigcirc$ Yes \qquad $\bigcirc$ No \qquad $\bigcirc$ Partially

% \begin{itemize}
%     \item[$\bigcirc$] Yes
%     \item[$\bigcirc$] No
%     \item[$\bigcirc$] Partially
% \end{itemize}

% \begin{table}[h]
% \begin{tabular}{lcr}
% $\bigcirc$ Yes & $\bigcirc$ No & $\bigcirc$ Partially
% \end{tabular}
% \end{table}

\noindent Having a correct understanding is the first and also the key step to coming up with a correct answer. Therefore, we are interested in investigating the capability of \gpt{} on question understanding.

\noindent \textbf{Q4.} \emph{Do you think the answer fully addresses the question? (Required)}

\noindent $\bigcirc$ Yes \qquad $\bigcirc$ No \qquad $\bigcirc$ Partially

% \begin{itemize}
%     \item[$\bigcirc$] Yes
%     \item[$\bigcirc$] No
%     \item[$\bigcirc$] Partially
% \end{itemize}

\noindent Certainly, the ultimate goal of SQA is to fully address the technical questions. Thus, we evaluate the capability which requires not only understanding the question but also providing the correct solutions with essential explanation.

\noindent \textbf{Q5.} \emph{Is there any factual error in the answer?  (Required)}

% \begin{itemize}
%     \item[$\bigcirc$] Yes
%     \item[$\bigcirc$] No
% \end{itemize}

\noindent $\bigcirc$ Yes \qquad $\bigcirc$ No

\noindent \emph{Hallucination} has been complained about by \gpt{} users and the issue has also been admitted by the OpenAI technical report~\cite{openaihallucination}.

\noindent \textbf{Q6.} \emph{If your answer to the previous question is Yes, please explain.}

\noindent \textbf{Q7.} \emph{Can you guess which one is generated by AI? (Required)}

% \begin{itemize}
%     \item[$\bigcirc$] Answer \#1.
%     \item[$\bigcirc$] Answer \#2.
%     \item[$\bigcirc$] I cannot recognize.
% \end{itemize}

\noindent $\bigcirc$ Answer \#1 \qquad $\bigcirc$ Answer \#2 \qquad $\bigcirc$ I cannot recognize.

\noindent Whether human users can distinguish \gpt{}-generated content is the key to embracing generative AI techniques.
In our case, we are interested in investigating whether humans can easily distinguish human and \gpt{}-generated answers. Thus, after participants answer the above questions for each answer, we ask them to guess which answer is more likely generated by \gpt{}.

\subsection{Result and Analysis}

% % Please add the following required packages to your document preamble:
% % \usepackage{graphicx}
% \begin{table}[h]
% % \resizebox{\columnwidth}{!}{%
% \begin{tabular}{cccc}
% \hline
%        & \begin{tabular}[c]{@{}c@{}}\# Unique\\ Questions\end{tabular} & \begin{tabular}[c]{@{}l@{}}\# Refused\\ Responses\end{tabular} & \begin{tabular}[c]{@{}c@{}}\# Valid\\ Responses\end{tabular}  \\ \hline
% Python & 20                                                            & 10                                                             & 30                                                                     \\
% Java   & 20                                                            & 4                                                              & 36                                                                     \\ \hline
% \end{tabular}%
% % }
% \caption{Statistic of Responses}
% \label{tab:responses_stats}
% \end{table}

% As shown in Table~\ref{tab:responses_stats},
We sampled 20 questions for each of the considered programming languages, i.e., 40 questions are selected in total. And for each question, we assign 2 participants to evaluate the quality of human and \gpt{} generated answers. In total, we create 40 questionnaires. 14 (17\%) of the questionnaires are not answered by participants since they think they are not knowledgeable to evaluate the answers to the specific questions. Following are the result and analysis based on the valid responses.

% \noindent \textbf{Q1.} \emph{How satisfied were you with the answer?}

% \noindent \textbf{Q2.} \emph{Please explain your rate on the answer.}

\begin{figure}[]
    \centering    \includegraphics[width=\columnwidth]{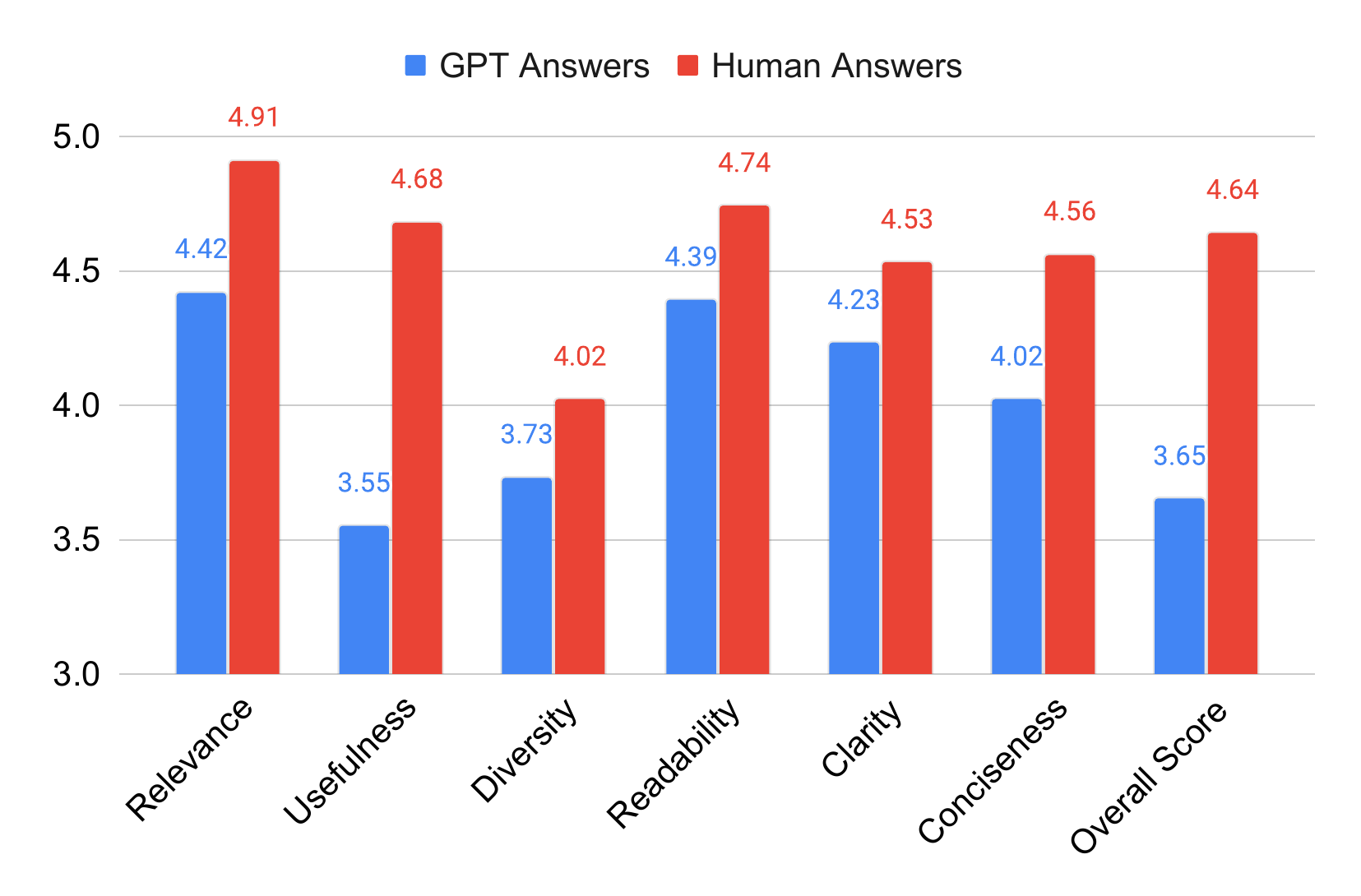}
    % \vspace{-7mm}
    \caption{Result of Q1. \emph{How satisfied were you with the answer?}}
    \label{fig:q1_res}
    \vspace{-5mm}
\end{figure}

% \usepackage{multirow}
% \usepackage{graphicx}
\begin{table*}[h!]
\centering
\caption{Results of Q3, Q4, and Q5}
\label{tab:rq-results}
\resizebox{.8\textwidth}{!}{%
\begin{tabular}{lcccccc}
\toprule
\multirow{2}{*}{Question}                                       & \multicolumn{3}{c}{Human-written Answers} & \multicolumn{3}{c}{\gpt{}-generated Answers} \\ \cline{2-7} 
                                                                & Yes      & Partially    & No      & Yes     & Partially    & No     \\ \hline
Q3. Correctly understands the question? & 95\%     & 3\%          & 2\%     & 80\%    & 11\%         & 9\%    \\ \hline
Q4. Fully addresses the question?       & 84\%     & 2\%          & 14\%    & 52\%    & 21\%         & 27\%   \\ \hline
Q5. Any factual error?                   & 2\%      & NA           & 98\%    & 27\%    & NA           & 73\% \\ \bottomrule
\end{tabular}%
}
\vspace{-4mm}
\end{table*}

% \begin{figure}[h]
%      \centering
%      \begin{subfigure}[b]{0.4\textwidth}
%          \centering
%     \includegraphics[width=\textwidth]{figs/q3-gpt-all.pdf}
%          \caption{\gpt{} Answers}
%          \label{fig:q3_gpt_all}
%      \end{subfigure}
%      \hfill
%      \begin{subfigure}[b]{0.4\textwidth}
%          \centering
%     \includegraphics[width=\textwidth]{figs/q3-human-all.pdf}
%          \caption{Human Answers}
%          \label{fig:q3_human_all}
%      \end{subfigure}
%         \caption{Results of Q3. \emph{Do you think the answer correctly understands the question?}}
%         \label{fig:q3_res}
% \end{figure}

\noindent \textbf{Q1.} \emph{How satisfied are you with the answer? (Required)}

\noindent \textbf{Q2.} \emph{Please explain your rate on the answer.}

% Q1 and Q2 are bundle questions.
Figure~\ref{fig:q1_res} presents the result of Q1. Among the 6 aspects of answer quality measurement,
human-written answers are consistently better than \gpt{}-generated answers. However, the gaps for the aspects, Relevance, Readability, and Clarity, are relatively small.
54\% of responses show human-written answers are better than \gpt{}-generated answers and only 11\% of them show \gpt{}-generated answers are better in terms of the overall score. \textbf{On average, human-written answers outperform \gpt{}-generated answers by 10\% on the overall score}.
From explanations received in Q2, we identify 2 additional reasons why participants scored human-written answers higher than \gpt{} ones:
% Shorterning suggestion
(1) \textbf{Generalizability}.
ChatGPT fails to generate appropriate answers for questions that do not appear in its training data, which is up to September 2021~\cite{chatgptoutdated}.
For instance, a question with ID 72166259 demonstrated this limitation: the human-written answer was preferred because \emph{``This answer solves the question with the latest knowledge''}.
% while the \gpt{}-generated answer, \emph{``This answer solves the problem using common approach which is not specific to the problem''}.
(2) \textbf{Correct-but-useless content}. We find that \gpt{}-generated answers may carry correct-yet-useless information. For example, one participant mentions that \emph{``... although the answer claims it can solve the problem, its content is not even relevant to the problem''}.
(3) \noindent\textbf{Make naive mistakes}.
Many works have demonstrated that generative AI has a big potential for software tasks, such as program repair \cite{pearce2022examining,xia2023automated,xia2022less}.
Based on our results, we find that the patch or code change suggestions generated by \gpt{} may carry simple errors that can be easily identified by humans. For example, one participant mentions that \emph{``... The code given seems to just repeat the code given in the question body with some minor modifications. Arrays::stream is not applicable to the stream of char[], so the code given can not be compiled and thus can not solve the question''}.

% Although the quantitative results show that \gpt{}-generated answers are of lower quality, we also find some counterexamples.
% \bowen{TBA}

% \noindent \textbf{Q3.} \emph{Do you think the answer correctly understands the question?}

\noindent \textbf{Q3.} \emph{Do you think the answer correctly understands the question? (Required)}

From Table~\ref{tab:rq-results}, we find that the number of human-written answers which fully understand the question is 15\% more than the number of \gpt{}-generated answers. However, there are still 80\% of \gpt{}-generated answers fully understood the question. It indicates that \textbf{\gpt{} has achieved a desirable capability on question understanding, although still not as good as humans.}.

% \noindent \textbf{Q4.} \emph{Do you think the answer fully addresses the question?}

% \begin{figure}[h]
%      \centering
%      \begin{subfigure}[b]{0.4\textwidth}
%          \centering
%     \includegraphics[width=\textwidth]{figs/q4-gpt-all.pdf}
%          \caption{\gpt{} Answers}
%          \label{fig:q4_gpt_all}
%      \end{subfigure}
%      \hfill
%      \begin{subfigure}[b]{0.4\textwidth}
%          \centering
%     \includegraphics[width=\textwidth]{figs/q4-human-all.pdf}
%          \caption{Human Answers}
%          \label{fig:q4_human_all}
%      \end{subfigure}
%         \caption{Results of Q4. \emph{Do you think the answer fully addresses the question?}}
%         \label{fig:q4_res}
% \end{figure}

\noindent \textbf{Q4.} \emph{Do you think the answer fully addresses the question? (Required)}

From Table~\ref{tab:rq-results}, we find that human-written answers are better than \gpt{}-generated answers in fully addressing the questions by a large margin (32\%).

% \noindent \textbf{Q5.} \emph{Is there any factual error in the answer?}

% \noindent \textbf{Q6.} \emph{If your answer to the previous question is Yes, please explain.}

% \begin{figure}[h]
%      \centering
%      \begin{subfigure}[b]{0.4\textwidth}
%          \centering
%     \includegraphics[width=\textwidth]{figs/q5-gpt-all.pdf}
%          \caption{\gpt{} Answers}
%          \label{fig:q5_gpt_all}
%      \end{subfigure}
%      \hfill
%      \begin{subfigure}[b]{0.4\textwidth}
%          \centering
%     \includegraphics[width=\textwidth]{figs/q5-human-all.pdf}
%          \caption{Human Answers}
%          \label{fig:q5_human_all}
%      \end{subfigure}
%         \caption{Results of Q5. \emph{Is there any factual error in the answer?}}
%         \label{fig:q5_res}
% \end{figure}

\noindent \textbf{Q5.} \emph{Is there any factual error in the answer?  (Required)}

\noindent \textbf{Q6.} \emph{If your answer to the previous question is Yes, please explain.}

% Q5 and Q6 are bundle questions.
From Table~\ref{tab:rq-results}, we find that human-written answers significantly carry fewer factual errors as compared to \gpt{}-generated answers.
For example, there is one factual error identified in the \gpt{}-generated answer that says ``\emph{This answer says that com.sun.xml.bind:jaxb-impl is no longer being actively developed or maintained, but the fact is that it is still getting new releases. See \url{https://mvnrepository.com/artifact/com.sun.xml.bind/jaxb-impl}}.''

% \noindent \textbf{Q7.} \emph{Can you guess which one is generated by AI?}

% \begin{figure}[h]
%     \centering    \includegraphics[width=.8\columnwidth]{figs/q7-all.pdf}
%     \caption{Result of Q7. \emph{Can you guess which one is generated by AI?}}
%     \label{fig:q7_res}
% \end{figure}

% Figure~\ref{fig:q7_res} presents the result of Q7.

\noindent \textbf{Q7.} \emph{Can you guess which one is generated by AI? (Required)}

We find that for 86\% of questions, participants can correctly distinguish human and \gpt{}-generated answers. Their further explanation provides the reason. For example, \emph{``\gpt{}-generated answers without any emotion''} and \emph{``From my experiences with ChatGPT, it usually provides complete and high-level answers with many redundant details... human-written answers are usually short and directly point out the solution''}.

%% file: 5_Threats.tex
% There are several threats that may potentially affect the validity of our study.
The programming languages considered in our experiments is a threat to external validity.
Similar to some prior works (e.g.,~\cite{yang2022answer}), we mitigate this threat by considering two popular programming languages, i.e., Java and Python.
Nevertheless, replicating our work for other programming languages is required to broaden our understanding of the capability of \gpt{}.
% A threat to internal validity related to our questionnaire design.
% We design the questions with pre-defined options related to the aspects (such as \emph{hallucination} issue) of \gpt{} which are heatedly discussed by OpenAI users.
% To mitigate biases led by these pre-defined options, we always encouraged our participants to use their own words to explain their answers.
Threats to construct validity are related to the used metrics. In this work, we use multiple automatic metrics to show the characteristic of \gpt{}-generated answers. However, considering automatic metrics alone may not be sufficient, we further mitigate this threat by performing both automatic and manual comparisons between human-written and \gpt{}-generated answers.

%% file: 7_related_work.tex
An et al.~\cite{an2017} conducted a case study on whether developers potentially reused code from Stack Overflow. Their studies suggested that developers may have copied the code of Android apps to answer Stack Overflow questions.
% Wu et al.~\cite{Wu2018} show that codes from Stack Overflow answers are often used as references or reused directly by other developers.
Fischer et al.~\cite{Fischer2017} and Zhang et al.~\cite{zhang2022} show these implications when reusing vulnerable code snippets from Stack Overflow. They highlight the security implications that can arise when developers reuse code from the platform without verifying them against potential vulnerabilities. 
Our study complements these studies by showing how generative AI, specifically ChatGPT can perform SQA support and how it compares to human-written answers as well as investigating the quality of their answers in various aspects linked to the aforementioned issues including correctness, usefulness, etc. 
% Olmar et al.~\cite{olmar2020} further studies how developpers discuss software reuse on Stack Overflow, 
% \cite{aoun2021} provides an empirical study on Stack Exchange Forum and Github Issues and provides insights on which issues is discussed on these forums. 
% \noindent\textbf{Automation of Q\&A}
% Gao et al.~\cite{Gao2020GeneratingQT} and ~\cite{liu2022sotitle} propose methods for automatic question title generation for Stack Overflow posts, leveraging LSTM networks with attention and pre-trained language models, respectively.
% Xu et al.~\cite{xu2017answerbot} present AnswerBot, which uses a ranking mechanism to retrieve the most important paragraph as a summary. Cao et al.~\cite{Cao2021} introduce DeepAns, a deep learning-based method for answer recommendation.
% Kou et al.~\cite{Kou2022, Kou2023} present a dataset for Stack Overflow post summarization and build ASSORT, a model that uses a pre-trained language model for automatic answer summarization.
% These works collectively demonstrate the potential of AI in automating various aspects of Q\&A, from question generation to answer summarization, while our study provides insights into the strength and limitations of automated AI-based SQA.

%% file: 8_Conclusion.tex
In this work, we compare answers generated by ChatGPT and humans in StackOverflow. Our study uncover the following findings: ChatGPT-generated answers are of lower quality than humans for all aspects considered. Still, they are promising and the gap is smaller for diversity, readability and clarity.
Our preliminary study sheds light on the potential future of SQA by summarizing the limitations of \gpt{}.
% In the future, we plan to propose a novel solution to mitigate the highlighted issues accordingly.

\vspace{0.2cm}\noindent{\bf Acknowledgement.} This research / project is supported by the National Research Foundation, Singapore, under its Industry Alignment Fund – Pre-positioning (IAF-PP) Funding Initiative. Any opinions, findings and conclusions or recommendations expressed in this material are those of the author(s) and do not reflect the views of National Research Foundation, Singapore. 

%% file: 0_main.bbl
\begin{thebibliography}{10}
\providecommand{\url}[1]{#1}
\csname url@samestyle\endcsname
\providecommand{\newblock}{\relax}
\providecommand{\bibinfo}[2]{#2}
\providecommand{\BIBentrySTDinterwordspacing}{\spaceskip=0pt\relax}
\providecommand{\BIBentryALTinterwordstretchfactor}{4}
\providecommand{\BIBentryALTinterwordspacing}{\spaceskip=\fontdimen2\font plus
\BIBentryALTinterwordstretchfactor\fontdimen3\font minus
  \fontdimen4\font\relax}
\providecommand{\BIBforeignlanguage}[2]{{%
\expandafter\ifx\csname l@#1\endcsname\relax
\typeout{** WARNING: IEEEtran.bst: No hyphenation pattern has been}%
\typeout{** loaded for the language `#1'. Using the pattern for}%
\typeout{** the default language instead.}%
\else
\language=\csname l@#1\endcsname
\fi
#2}}
\providecommand{\BIBdecl}{\relax}
\BIBdecl

\bibitem{chatgpt}
``Chatgpt,'' \url{https://openai.com/blog/chatgpt}.

\bibitem{chatgptuser}
``Chatgpt users,''
  \url{https://www.reuters.com/technology/chatgpt-sets-record-fastest-growing-user-base-analyst-note-2023-02-01/}.

\bibitem{chatgpt4sqa}
``Chatgpt for software q\&a,''
  \url{https://www.infoworld.com/article/3689172/chatgpt-and-software-development.html}.

\bibitem{sobanchatgpt}
``Stack overflow banned chatgpt,''
  \url{https://meta.stackoverflow.com/questions/421831/temporary-policy-chatgpt-is-banned}.

\bibitem{sotraffic}
``Stack overflow traffic,''
  \url{https://stackoverflow.blog/2018/04/26/stack-overflow-isnt-very-welcoming-its-time-for-that-to-change/}.

\bibitem{sodecline}
``Stack overflow traffic decline,''
  \url{https://www.similarweb.com/website/stackoverflow.com/#demographics}.

\bibitem{calefato2018ask}
F.~Calefato, F.~Lanubile, and N.~Novielli, ``How to ask for technical help?
  evidence-based guidelines for writing questions on stack overflow,''
  \emph{Information and Software Technology}, vol.~94, pp. 186--207, 2018.

\bibitem{ahasanuzzaman2016mining}
M.~Ahasanuzzaman, M.~Asaduzzaman, C.~K. Roy, and K.~A. Schneider, ``Mining
  duplicate questions in stack overflow,'' in \emph{Proceedings of the 13th
  International Conference on Mining Software Repositories}, 2016, pp.
  402--412.

\bibitem{baltadzhieva2015predicting}
A.~Baltadzhieva and G.~Chrupa{\l}a, ``Predicting the quality of questions on
  stackoverflow,'' in \emph{Proceedings of the international conference recent
  advances in natural language processing}, 2015, pp. 32--40.

\bibitem{soacceptedanswer}
``Stack overflow accepted answers,''
  \url{https://stackoverflow.com/help/accepted-answer}.

\bibitem{openaitextsim}
``Openai text similarity,''
  \url{https://platform.openai.com/docs/api-reference/chat/create}.

\bibitem{openaitokenizer}
``Openai tokenizer,'' \url{https://platform.openai.com/tokenizer}.

\bibitem{openaiembeddings}
``Openai text and code embeddings,''
  \url{https://openai.com/blog/introducing-text-and-code-embeddings}.

\bibitem{cohen1960coefficient}
J.~Cohen, ``A coefficient of agreement for nominal scales,'' \emph{Educational
  and psychological measurement}, vol.~20, no.~1, pp. 37--46, 1960.

\bibitem{xu2020reinventing}
B.~Xu, L.~An, F.~Thung, F.~Khomh, and D.~Lo, ``Why reinventing the wheels? an
  empirical study on library reuse and re-implementation,'' \emph{Empirical
  Software Engineering}, vol.~25, pp. 755--789, 2020.

\bibitem{xu2017answerbot}
B.~Xu, Z.~Xing, X.~Xia, and D.~Lo, ``Answerbot: Automated generation of answer
  summary to developers' technical questions,'' in \emph{2017 32nd IEEE/ACM
  international conference on automated software engineering (ASE)}.\hskip 1em
  plus 0.5em minus 0.4em\relax IEEE, 2017, pp. 706--716.

\bibitem{yang2022answer}
C.~Yang, B.~Xu, F.~Thung, Y.~Shi, T.~Zhang, Z.~Yang, X.~Zhou, J.~Shi, J.~He,
  D.~Han \emph{et~al.}, ``Answer summarization for technical queries: Benchmark
  and new approach,'' in \emph{37th IEEE/ACM International Conference on
  Automated Software Engineering}, 2022, pp. 1--13.

\bibitem{openaihallucination}
``Hallucination issue of chatgpt,''
  \url{https://cdn.openai.com/papers/gpt-4.pdf}.

\bibitem{chatgptoutdated}
``Training data of chatgpt,''
  \url{https://help.openai.com/en/articles/6783457-what-is-chatgpt}.

\bibitem{pearce2022examining}
H.~Pearce, B.~Tan, B.~Ahmad, R.~Karri, and B.~Dolan-Gavitt, ``Examining
  zero-shot vulnerability repair with large language models,'' in \emph{2023
  IEEE Symposium on Security and Privacy (SP)}.\hskip 1em plus 0.5em minus
  0.4em\relax IEEE Computer Society, 2022, pp. 1--18.

\bibitem{xia2023automated}
C.~S. Xia, Y.~Wei, and L.~Zhang, ``Automated program repair in the era of large
  pre-trained language models,'' in \emph{Proceedings of the 45th International
  Conference on Software Engineering (ICSE 2023). Association for Computing
  Machinery}, 2023.

\bibitem{xia2022less}
C.~S. Xia and L.~Zhang, ``Less training, more repairing please: revisiting
  automated program repair via zero-shot learning,'' in \emph{Proceedings of
  the 30th ACM Joint European Software Engineering Conference and Symposium on
  the Foundations of Software Engineering}, 2022, pp. 959--971.

\bibitem{an2017}
L.~An, O.~Mlouki, F.~Khomh, and G.~Antoniol, ``Stack overflow: A code
  laundering platform?'' in \emph{2017 IEEE 24th International Conference on
  Software Analysis, Evolution and Reengineering (SANER)}, 2017, pp. 283--293.

\bibitem{Fischer2017}
F.~Fischer, K.~Böttinger, H.~Xiao, C.~Stransky, Y.~Acar, M.~Backes, and
  S.~Fahl, ``Stack overflow considered harmful? the impact of copy\&paste on
  android application security,'' in \emph{2017 IEEE Symposium on Security and
  Privacy (SP)}, 2017, pp. 121--136.

\bibitem{zhang2022}
H.~Zhang, S.~Wang, H.~Li, T.-H. Chen, and A.~E. Hassan, ``A study of c/c++ code
  weaknesses on stack overflow,'' \emph{IEEE Transactions on Software
  Engineering}, vol.~48, no.~7, pp. 2359--2375, 2022.

\end{thebibliography}
